# Observation of Toroidal Pulses of Light

A. Zdagkas[1], Y. Shen[1], N. Papasimakis[1], N. I. Zheludev[1,2]

[1]Optoelectronics Research Centre & Centre for Photonic Metamaterials, University of Southampton, Highfield SO17 1BJ, UK

[2]Centre for Disruptive Photonic Technologies, School of Physical and Mathematical Sciences and the Photonics Institute, Nanyang Technological University, 637371, Singapore

C. McDonnell[3], J. Deng[4,5], G. Li[4,5], T. Ellenbogen[3]

[3]Department of Physical Electronics, School of Electrical Engineering, Tel-Aviv University, 6997801 Tel Aviv, Israel

[4]Department of Materials Science and Engineering, Southern University of Science and Technology, 518055, Shenzhen, China

[5]Shenzhen Institute for Quantum Science and Engineering, Southern University of Science and Technology, 518055, Shenzhen, China

**Abstract**: The transverse electromagnetic waves are major information and energy carriers. In 1996, Hellwarth and Nouchi theoretically identified a radically different, non-transverse type of electromagnetic pulses of toroidal topology. These pulses, which are propagating counterparts of localized toroidal dipole excitations in matter and exhibit unique electromagnetic wave properties, have never been observed before. Here, we report the generation and characterization of such optical and terahertz Toroidal Light Pulses (TLPs), launched from tailored nanostructured metasurfaces comprising toroidal emitters. This achievement paves the way for experimental studies of energy and information transfer with TLPs, their space-time "entanglement", and their light-matter interactions involving anapoles, localized space-time entangled excitations, skyrmions, and toroidal qubits that are of growing interest for the fundamental science of light and applications.

The emerging research field of toroidal electrodynamics has been attracting growing interest since the experimental observation of the oscillating toroidal dipole in 2010 [1] and dynamic electromagnetic anapoles in 2013 [2]. This has been stimulated by the intriguing electromagnetic properties of toroidal dipoles, such as odd space and time symmetry, complementarity to the electric and magnetic dipoles [3], and predictions of their involvement in non-reciprocal electromagnetic interactions [4]. Dynamic toroidal dipoles have been observed in artificial materials and nanostructures in the microwave regime and across the optical part of the spectrum [5-9]. They are now recognized as indispensable contributors to the linear and nonlinear optical properties of matter [10,11]. As such, new types of spectroscopy, selective to optical transitions enabled by toroidal dipoles, are being developed based on solvatochromism [12] and standing wave illumination [13]. Furthermore, dynamic anapoles, co-located non-radiating configurations of oscillating electric and toroidal dipoles [2] now attract growing interest as a vehicle for high quality resonant devices [14-16], quantum qubits [17] and even reservoirs of dark energy [18]. Although anapole modes do not radiate electromagnetic fields, they arguably emit vector potential waves that are not removable by gauge transforms and may lead to new Aharonov-Bohm type phenomena [19,20].

Since 1996, it has been understood by the work by Hellwarth and Nouchi, that oscillating toroidal excitations can also exist in the form of bursts of electromagnetic energy propagating in free space at the speed of light [21]. Although such Toroidal Light Pulses (TLPs) travel at the speed of light, they are radically different from conventional plane electromagnetic waves on several counts:

(a) *TLPs possess non-transverse electromagnetic field components*. Indeed, the conventional plane electromagnetic waves that we know since the early days of Heinrich Rudolf Hertz and that are used in modern free-space telecommunication technologies are transverse electromagnetic waves, in which the electric and magnetic fields oscillate perpendicularly to the direction of propagation. Toroidal light pulses are not transverse, but instead present toroidal topology with the magnetic field tracing the body of a donut-like shape, while the electric field traces its surface exhibiting electric field components that are oriented along the direction of pulse propagation, see Fig. 1. A complementary form of TLPs with a component of magnetic field oscillating along the direction of propagation can be obtained by exchanging the electric and magnetic fields [21]. These two forms are known as transverse magnetic (TM) and transverse electric (TE) Toroidal Light Pulses, respectively.

(b) *TLPs are space-time non-separable*. Plane waves belong to a family of solutions to Maxwell's equations that can be presented as a product of time and space dependent complex exponential functions and includes, for example, Gaussian, Bessel, Airy waves and pulses. However, it has been known since the early 1980s that Maxwellian electrodynamics admit solutions, where the space and time dependences are non-separable, i.e. cannot be presented as a product of the time only and space only dependent terms [22-26]. TLPs belong to this category: they are exact solutions to Maxwell's equations, but they are of non-separable nature [21]. The space-time non-separability of the TLPs manifests in the space-spectrum domain as a position-dependent frequency content across the transverse plane. Here, the lower frequency components dominate at the periphery of the toroidal pulse, while higher frequencies prevail at its central area (see Fig. 1A-B).

(c) *TLPs are space-time energy localizations.* Elementary TLPs exist as short localized, single and 1 ½ cycle bursts of radiation with broad spectrum and finite total energy. Upon propagation, the TLPs focus and defocus and experience reshaping due to Gouy phase shift: they evolve from 1 ½ cycle (single-cycle) away from focus to single-cycle (1 ½ cycle) pulse at focus and back to 1 ½ cycle (single-cycle) pulse, see Fig. 1A. The Hellwarth & Nouchi Toroidal Light Pulse is characterized by two parameters, $q_1$ and $q_2$, which correspond to an effective wavelength and length of the focal region, respectively [21].

(d) *TLPs exhibit a space-time toroidal topology.* The unique space-time toroidal structure of TLPs contains multiple singularities and areas of energy backflow (zones where electromagnetic energy travels in a direction opposite to the direction of pulse propagation) that are identifiable at the low-intensity areas of the pulse [27].

To date, the toroidal light pulses have not been observed since their generation presents a formidable challenge due to their topology and non-separable space-time structure. Here we report two complementary approaches to the generation of such pulses, by using nanostructured emitters of toroidal topology that are driven by conventional laser light pulses (see Fig. 1C&D). In the optical part of the spectrum, we use a linear metasurface designed to scatter conventional laser pulses into few-cycle optical TLPs (Fig. 1C), while in the terahertz range we employ a nonlinear metasurface that rectifies the envelopes of incident laser pulses and re-radiate them as single-cycle terahertz TPLs (Fig. 1D).

The generation scheme for optical TLPs comprises a linear-to-radial polarization converter and a nanostructured metasurface (see Fig. 1C) that are driven by few-cycle (<10 fs at FWHM) linearly polarized laser pulses with central wavelength of ~800 nm. The metasurface consists of concentric gold rings: the width of the rings and thus their resonant plasmonic properties vary with the ring radius in such a way that re-radiated light accrues phases and amplitudes according to the spatiotemporal coupling of TLPs.

In our experiments, we targeted the generation of ideal TLPs that can be fully defined by the effective wavelength, $q_1$, and Rayleigh range, $q_2$. The effective wavelength is determined by the center wavelength of our laser system (~800 nm). The structural dimensions of the metasuface and divergence of the pump beam entering it control the Rayleigh range. The plasmonic metasurface is designed with the transmission resonance wavelength varying along the radial direction to achieve $q_1$=192 nm and $q_2$=75,000$q_1$. To emphasize the importance of the design, a second metasurface design was used to create pulses of toroidal symmetry, but excessively large space-spectrum coupling incompatible with the spatiospectral structure of ideal TLPs.

The spatiotemporal and spatiospectral structure of the generated TLPs were characterized by hyperspectral imaging of their transverse profiles (as constructed from multiple images at different wavelengths) and by spatially resolved Fourier transform interferometry [28-29], allowing the retrieval of spectral amplitude and phase for all polarization components of the electric field at any point of the TLP.

The results of spectral and interferometric measurements confirm the toroidal topology of the generated pulses and compliance with the characteristics of the ideal TLP (see Fig. 1A). In the generated pulses, the lower frequency components are dominant at the periphery of the profile, while higher frequencies prevail in its central part (Fig. 2B&E). Here, the main qualitative measure of the correct space-spectrum non-separability is the gradual wavelength dependent shift of the radial position of the spectral components of the experimentally generated pulses, that closely match that of the ideal TLP (Fig. 2A&D). In comparison, the hyperspectral profile of the pulses generated by the metasurface with excessive space-spectrum coupling (Fig. 2C&F) deviates significantly from that of the ideal TLP. The vectorial spatial interferometry used in our experiments reconstructs all the components of the electric and magnetic fields. Figure 2G-I shows isosurfaces of the electric field of the generated TLP at a level of 60% of its maximum. Red and blue colors correspond to the two half-cycles of the pulse. As expected, the transverse electric field vanishes at the center of the pulse (Fig. 2H) and exhibit substantial longitudinal components (see Fig. 2I).

To quantify the similarity of the targeted pulses with the ideal TLPs, we employ a state tomography approach analyzing a discrete set of spatial and spectral states of the pulses [26]. This method returns a fidelity measure, 0<F<1, where F=1 indicates perfect similarity of the pulses, while small fidelity values correspond to lack of similarity. The pulses generated with the optimized metasurface (Fig. 2B&E) exhibited high fidelity of F=0.79, while pulses generated by the metasurface with excessively large spectral shift (Fig. 2C&F) delivered F=0.1.

Therefore, spatiotemporal and spatiospectral characterization of the generated optical pulses reveals the main features of ideal Toroidal Light Pulses: a) The presence of non-transverse electric field component (see Fig. 2I); b) Nonseparable spatiospectral structure of the generated pulses as quantified by high fidelity value; c) Space-time energy localization. Owing to the limited bandwidth of the laser system, the generated pulses are of few-cycles duration and thus are not elementary single cycle TLPs, generation of which will be described in the THz section of this paper; d) The generated optical pulses exhibit a profound toroidal topology; the observation of singularities and energy backflow features were not targeted in the current work as they occur at the areas of lower energy inaccessible to vectorial spatial interferometry.

We generated single cycle TLPs in the terahertz part of the spectrum through optical rectification of femtosecond near-infrared pulses (~50 fs with center wavelength of ~1500 nm) on a Pancharatnam-Berry phase metasurface [30-35]. We used a plasmonic metasurface consisting of a cylindrically symmetric array of gold Y-shaped meta-atoms with three-fold rotational symmetry (see Fig. 1D). Orientation of the principal axis of the meta-atoms at an angle $\theta$ with respect to the linear polarization direction of the pump near-IR pulse results in the generation of single-cycle THz waves, linearly polarized at an angle $3\theta$ [35], as shown schematically in Fig. 1D. Arranging the meta-atoms on the circularly symmetric metasurface with their principal axis rotating 120° around the circumference can be used for the generation of TM polarized or TE polarized TLPs, depending on the incident polarization of the pump pulse (0° for TE and 45° for TM). The generated pulses are characterized in time domain with raster scanning in two dimensions. The generated THz pulse exhibits a sub 2-cycle duration as the 1 ½ - cycle ideal TLP, (see Fig. 3). Both the ideal TLP and the generated pulse exhibit opposite signs of $E_x$ field in opposite sides of the x=0 plane (red and blue colored regions in Fig. 3A&C) and take minimum values at the center of the pulse (for x=0). The same spatial characteristics are measured for the $E_y$ field in opposite sides of the y=0 plane, indicating polarization along the radial direction. The space-time non-separable nature of the ideal and generated TLPs is presented in Fig. 3B&D, respectively, in the form of the spatial profile of the corresponding frequency spectra. The space-spectral non-separability here manifests in the shift of the intensity maxima position from short to longer radii with decreasing frequency: the corresponding traces of the intensity maxima follow closely spaced trajectories in the spatiospectral plane (frequency-radius). The corresponding value of space-spectral non-separability fidelity is F=0.80. Therefore, the generated THz pulses exhibit the main features of the ideal TLPs: (a) Electric field with radial and longitudinal components (see Fig. 3C); (b) Space-time non-separability with high fidelity (see Fig. 3D); (c) Space-time localization and pulse duration close to the single- to 1 ½- cycle ideal TLP (Fig. 3A&C); (d) Toroidal topology.

In conclusion, we have demonstrated for the first time the generation and detection of TLPs in the optical and THz parts of the spectrum and mapped their temporal and spectral characteristics and space-time structure. Such light pulses are prime candidates for the investigation of toroidal light-matter interactions, in particular with respect to the excitation of toroidal and anapole modes in matter. Their space-time non-separable structure is expected to lead to non-trivial light-matter interactions and raises the question of violations of microscopic reciprocity [4]. Finally, their unique propagation dynamics will lead to novel spectroscopic applications and exotic information and energy transfer schemes.

**Acknowledgments:** This work was supported by the UK Engineering and Physical Sciences Research Council (Grant No. EP/ M009122/1), the MOE Singapore (Grant No. MOE2016-T3- 1-006), the European Research Council (Advanced grant FLEET-786851, Funder Id: http://dx.doi.org/10.13039/501100000781), and the Defense Advanced Research Projects Agency (DARPA) under the Nascent Light Matter Interactions program. T.E. acknowledges funding from the European Research Council (ERC) under the European Union's Horizon 2020 research and innovation program (Grant Agreement No. 715362). G. L. is financially supported by National Natural Science Foundation of China (91950114 and 11774145), Guangdong Provincial Innovation and Entrepreneurship Project (2017ZT07C071).

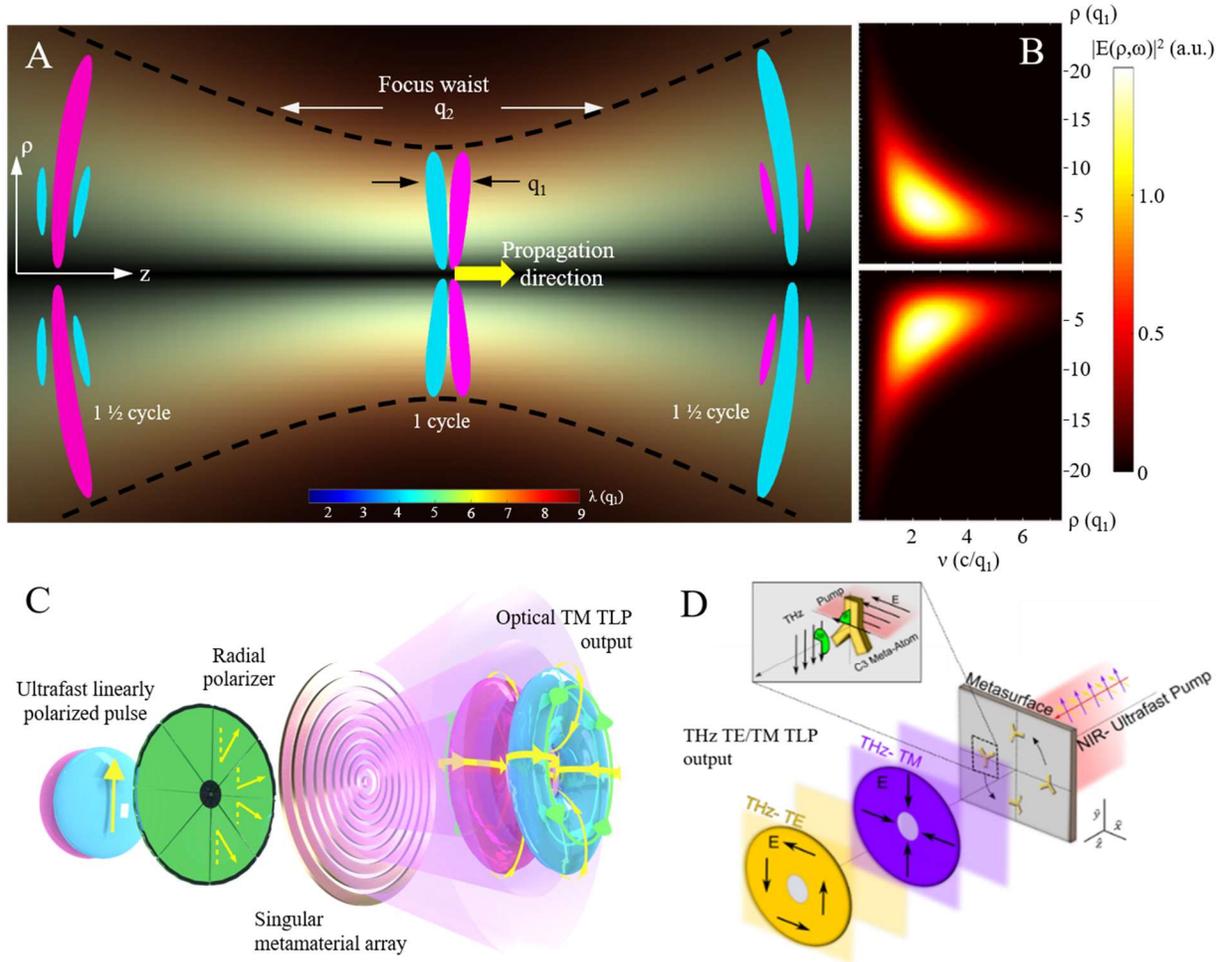

**Fig. 1. Characteristics of Toroidal Light Pulses and blueprint for their generation. (A-B)** Spatiotemporal and spatiospectral structure of the TLP. (A) Propagation of the toroidal light pulse. The frequency and intensity are represented by color and brightness correspondingly. The dashed lines mark the trajectory traced by the waist of the pulse. (B) Space-time coupling in the toroidal light pulse is manifested in the distribution of frequency components along the radial direction $\rho$: the higher frequency components are closer to the pulse center and lower frequency components are at the periphery of the pulse ($|\rho|>>q_1$). **(C)** The generation of TM TLPs in the optical part of the spectrum. A linearly polarized ultrafast pulse is converted to a radially polarized pulse by a segmented waveplate that acts as broadband polarization converter. The TLPs are then launched by a singular plasmonic metasurface excited with the radially polarized pulse. **(D)** Generation of THz, TM and TE, TLPs using plasmonic metasurfaces. The metasurface is illuminated with near-IR ultrashort pulses (~50 fs). The Y-shaped meta-atoms (see inset) then radiate electromagnetic waves in the THz spectral region through optical rectification. Their spatial arrangement is designed to generate the TLPs.

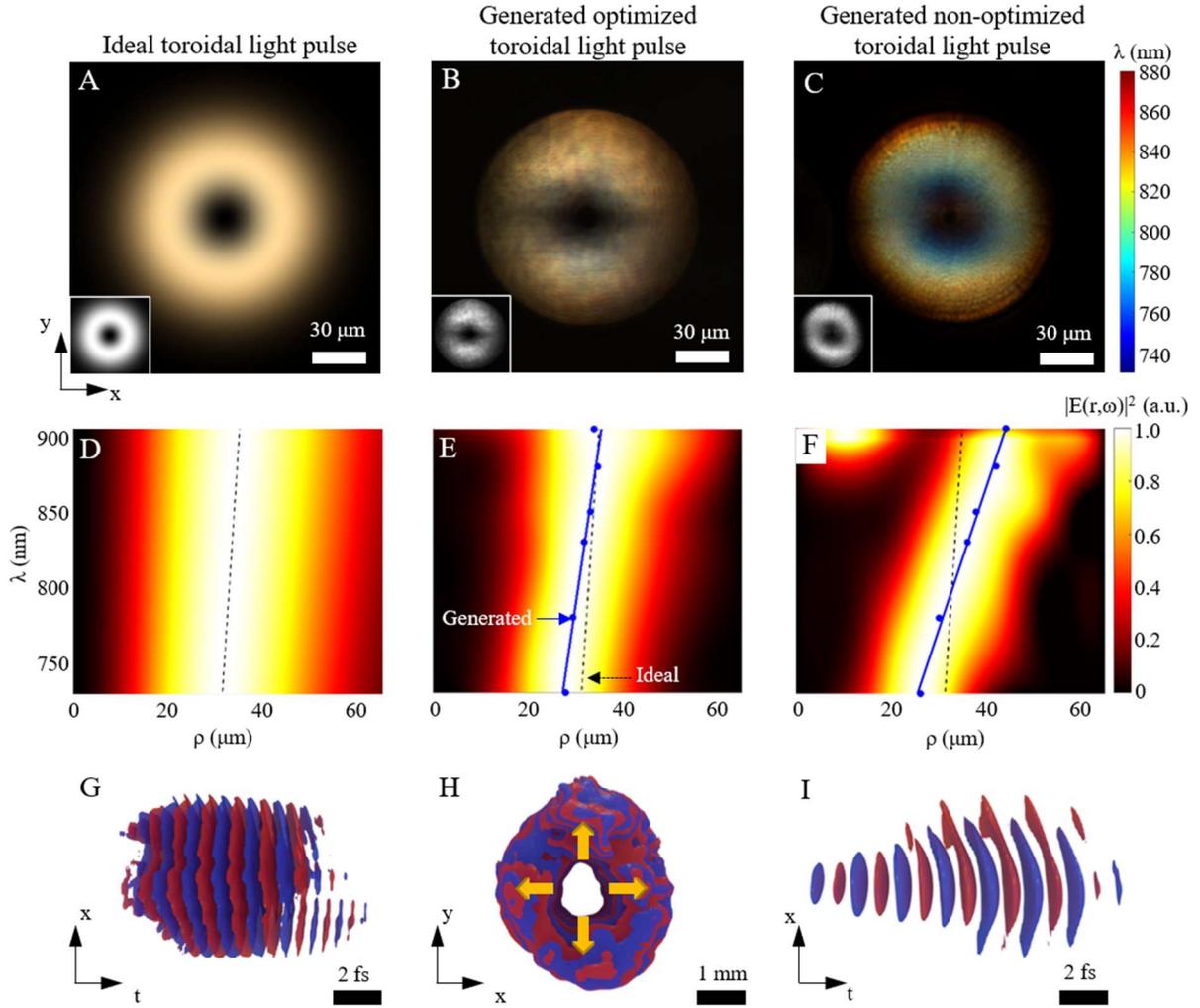

**Fig. 2. The spatiospectral and spatiotemporal structure of generated optical Toroidal Light Pulses.** **(A-C)** A comparison of hyperspectral images of the transverse profiles of ideal TLPs computed analytically (A) to the experimental profiles of experimentally generated TLPs (B). Panel (C) shows the profile of the pulse generated by a metasurface with excessively large space-spectrum coupling. Insets to (A-C) present the intensity profiles of the pulses shown in the corresponding panels. **(D-F)** Radial distribution of frequency components for the pulses presented in (A-C). The dashed and solid lines track the intensity maxima of the ideal TLP and experimentally generated pulses, correspondingly. **(G-I)** Details of the structure of the optimized pulse retrieved by interferometry. The interference is performed on the expanded and collimated pulse in contrast to panels (A-F) that correspond to the pulse at focus. Panels (G,H) show side and front views of the transverse electric field component. Panel (I) shows a side view of the longitudinal component.

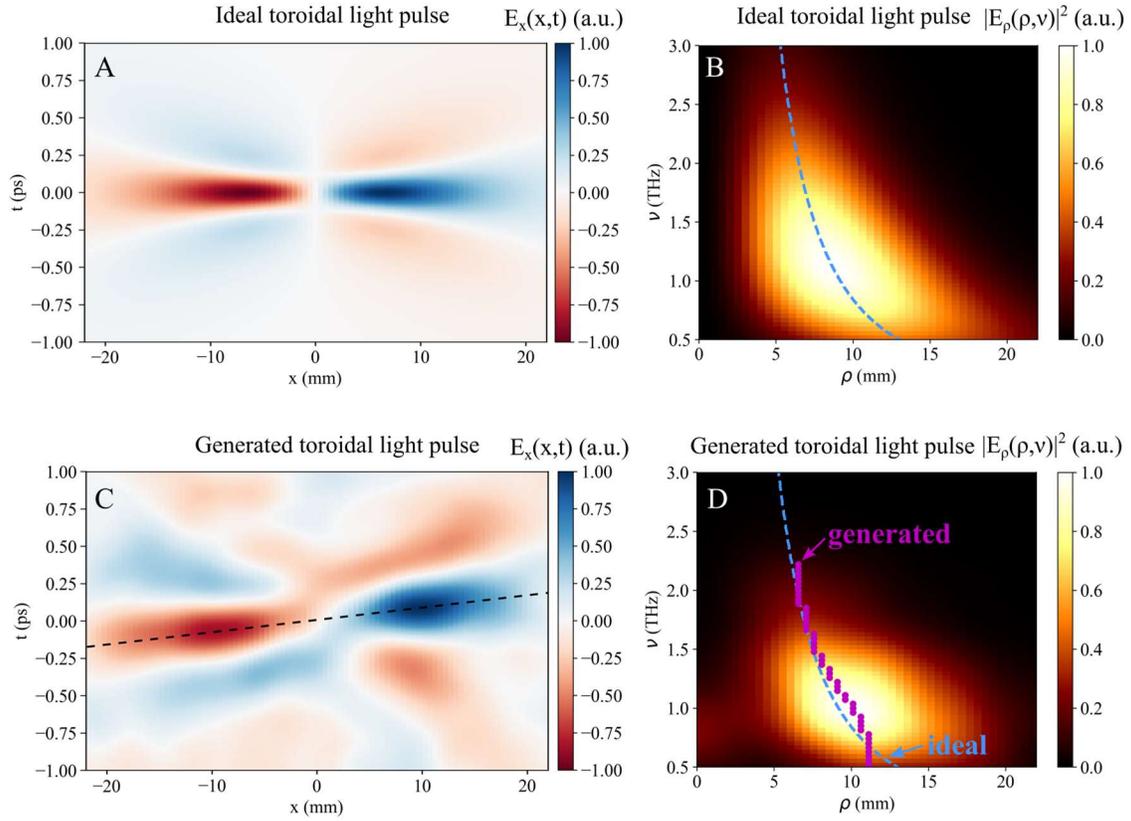

**Fig. 3. The spatiotemporal and spatiospectral structure of generated THz toroidal light pulses.** Spatiotemporal map (A,C) and spatial variations of the spectrum (B,D) for an ideal (A,B) and the generated (C,D) THz TM TLP. Blue dashed lines and purple markers in (B,D) track the intensity maxima of the ideal and experimentally generated pulses, respectively. The black dashed line in (C) serves as guide to the eye indicating a small tilt in the experimentally generated pulse. The parameters of the ideal TLP are $q_1=63\mu m$ and $q_2=56,000 q_1$.